\documentclass[twocolumn,amsmath,amssymb]{revtex4}
\usepackage{graphicx}
\usepackage{setspace}
\usepackage{epsfig}
\begin{document}
\title{Nonlinearity in Bacterial Population Dynamics:\\ Proposal for Experiments for the Observation of Abrupt Transitions in Patches}
\author{V. M. Kenkre and Niraj Kumar}
\address{Consortium of the Americas for Interdisciplinary Science and Department
         of Physics and Astronomy, University of New Mexico, Albuquerque, NM 87131, USA}
\begin{abstract}
An explicit proposal for experiments leading to abrupt transitions in spatially extended bacterial populations in a Petri dish is presented on the basis of an exact formula obtained through an analytic theory. The theory provides accurately the transition expressions in spite of the fact that the actual solutions, which involve strong nonlinearity, are inaccessible to it. The analytic expressions are verified through numerical solutions of the relevant nonlinear equation. The experimental set-up suggested uses opaque masks in a Petri dish bathed in ultraviolet radiation as in Lin et al., Biophys. J. {\bf 87}, 75 (2004) and Perry, J. R. Soc. Interface {\bf 2}, 379 (2005) but is based on the interplay of two
distances  
 the bacteria must traverse, one of them favorable and the other adverse.
As a result of this interplay feature, the experiments proposed introduce highly enhanced reliability in interpretation of observations and in the potential for extraction of system parameters.
\end{abstract}
\maketitle

Phenomena displaying abrupt transitions are of special interest to a variety of sciences, including physics
and biology. In some physical cases they arise from cooperative interactions
among a large number of constituents (e.g., molecules or spins), and in
others from nonlinearities in interaction inherent in the system, (e.g., in some mechanical systems with only a few degrees of freedom.) In the first
form they are known as phase transitions \cite{phase}, in the second as bifurcations \cite{bifurc}. Abrupt phenomena also command attention in the context of
extinction of populations, a subject of obvious interest to biology \cite{popu}. Transitions in populations, therefore, constitute an exciting
topic of interdisciplinary science combining physics and biology, and
the present paper reports a theory and proposes an experiment in this
topic. The special feature of the proposed experiment is that it may be performed with relatively simple equipment and
measurement techniques.

The system to be considered consists of bacteria in a Petri dish,
allowed to grow and move, spatial selection being imposed via lethal
ultraviolet radiation that is incident on the dish but punctuated by
opaque masks that protect the bacteria in chosen regions. Experiments
under such a setup were initiated several years ago by Lin et al. \cite{lin}
who used moving masks in response to a theoretical analysis \cite{nelson}
that focused on spatial disorder. A quite different experiment using stationary masks of varying sizes, and employing bacterial extinction as the key phenomenon was then proposed \cite{kk}. That proposal was carried out experimentally \cite{perry}
 and the predictions of ref. \cite{kk} were verified. Related investigations on this topic may be
found in \cite{other,mann,ballard}, the general problem of bacterial
dynamics in Petri dishes having been addressed in various papers and contexts
\cite{general}.

The underlying assumption behind most of these studies is that the
bacteria obey a simple Fisher equation \cite{fisher} for the time evolution of their
dynamics, their population density $u(x,t)$, where $x$ is the position in
a 1-dimensional space (the linear dimension of the Petri dish), and $t$
is the time, being governed by
\vspace{-0.02705cm}
\begin{equation}
\frac{\partial u(x)}{\partial t}=D\frac{\partial^2 u(x)}{\partial x^2}+au(x)-
bu^2(x).
\label{fisher}
\end{equation}
Here $a$ and $b$ are, respectively, the growth rate and a competition parameter arising
from the resources being limited, and $D$ is the bacterial diffusion constant.

Equation (\ref{fisher}) generally does not permit analytic solutions for arbitrary times but can be solved in the steady state explicitly in
terms of Jacobian elliptic functions, if the bacterial population is assumed to vanish at the
edges of a finite region. Physically, this could correspond to an idealization in which, while the mask protects the bacteria
from the lethal effects of the ultraviolet radiation, the latter is so potent
that bacterial population must vanish everywhere outside the mask
where the radiation impinges on the bacteria. Mathematically, these are Dirichlet boundary conditions. The elliptic function
solution, given independently in a recent analysis \cite{kk},
but known decades earlier through the work of Skellam \cite{skellam}, and elucidated in textbooks \cite{kot},
leads to the well-known KISS transition \cite{kiss}: the steady
state bacterial population vanishes for any width of the mask lower than a critical value that, interestingly, depends on the diffusion constant $D$ and the growth rate $a$, but not on the nonlinearity parameter $b$. The specific expression for the critical mask width is $ \pi \sqrt{D/a}$. Combining this expression with information about bacterial diffusion constants and growth rates discussed by Mann
in his Ph.D. thesis \cite{mann}, Kenkre and Kuperman calculated the critical mask width to be around $0.5 cm$, and suggested that an experiment be
carried out to observe the transition. Perry \cite{perry} followed the suggestion, and reported observing a transition width of 0.8 cm in his experiments on a non-chemotactic strain RP9535 of \emph{E. coli} bacteria.
Since the Dirichlet boundary condition is only an extreme idealization, Perry argued, appropriately, that it is preferable to invoke the analysis of Ludwig et al. \cite{ludwig} for
quantitative verification. That analysis 
 assumes that the growth rate $a$ is negative in the region outside the mask (and of magnitude $a_1$) rather than infinite as  would correspond to Dirichlet boundary conditions,  and arrives at the critical mask width
as being $L=2 \sqrt{D/a}\arctan \sqrt {a_1/a}$.
The Dirichlet expression is recovered from this Ludwig formula  if
the  effect of the ultraviolet radiation outside the mask is ``infinitely'' lethal.

\section{Proposal for Experiment}
\begin{figure}
 \centerline{\includegraphics[width=0.4\textwidth]{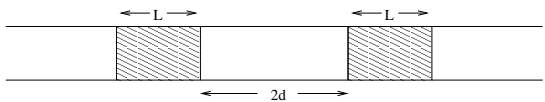}}
 \caption{Our proposed experimental set-up showing two favorable regions each of length $L$ separated
          by a hostile region of length $2d$. The focus of observation is the interdependence of the critical values of $L$ and $d$ at which extinction occurs.}
\label{fig1}
\end{figure}

Need for the new experimental proposal that we present here arises from the fact that the stationary single mask experiment \cite{kk,perry} uses for its interpretation a number of inputs which suffer from a certain degree of uncertainty. Some of these uncertainties are about the values of $D$ and $a$ \cite{mann}, and others about the extent to which other phenomena such as signaling \cite{perry} that occur in bacterial movement might affect the outcome of the experiment. Because absolute values of critical lengths may be difficult to obtain with an acceptable degree of accuracy, we propose here stationary mask observations, that focus, instead, on the \emph{interdependence of two critical lengths}. The essential feature of this present proposal is to
have more than one mask in the Petri dish, so that there are two
controllable lengths, a favorable length associated with a livable
region for the bacteria, and a hostile length associated with an
unlivable region. The double control possible in the proposed experiment opens the prediction space from a single
value to an infinity of values. The variation of the critical value of one of
the lengths when the other is changed presents an entire \emph{relationship} that is directly observable rather than
a single value and therefore could lead to a much cleaner and trustworthy interpretation of the observations.

The simplest system to study consists of two masks, each of length $L $, separated
by a distance of length $2d$ as shown in Fig. \ref{fig1}.
The previously investigated
case with a single mask \cite{kk,perry} corresponds, obviously, to infinite $d$. To test the idea behind the new proposal, we carried out numerical studies of the full nonlinear Fisher equation (\ref{fisher}) with given values of $D$, $a$, $a_1$, and $b$ (10, 0.1, -0.9 and 1 in appropriate units.) All lengths were expressed as ratios to the diffusion length $\sqrt{D/a}$. The numerical studies used an explicit finite differences scheme, $x$ being discretized  in intervals of 0.1. The convergence to a steady state was analyzed by measuring the distance between successive solutions.

\begin{figure}
  \centerline{\includegraphics[width=0.4\textwidth]{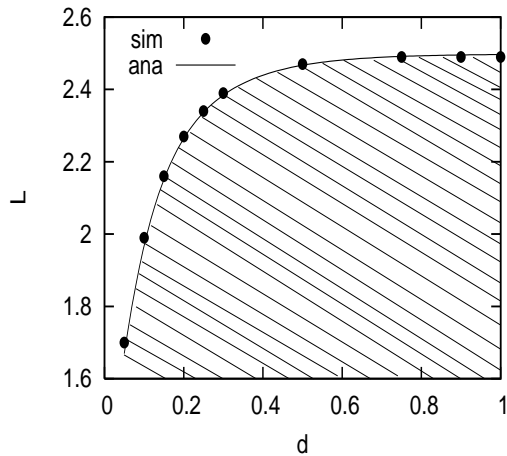}}
  \caption{Basis of the experimental proposal: Shown is the interdependence of the critical values of the favorable and hostile lengths $L$ and $d$ respectively. The former is the width of the mask and the $2d$ is the distance between the two masks (see Fig. 1), both lengths being expressed in this plot in units of the diffusion length $ \sqrt{D/a}$ where $a$ is the growth rate under the masks. Outside the masks $a$ is replaced by $-a_1$ to represent the harsh effect of the ultraviolet light, the magnitude of $a_1$ being taken for the purposes of this plot to be 9 times that of $a$. We solved Eq. \ref{fisher} numerically for various different values of the nonlinear parameter $b$ and found that the results did not depend on those values provided $b$ was non-zero and positive. The shaded region represents pairs of $L$ and $d$ values that lead to bacterial extinction in the steady state.
  Filled circles mark the onset of extinction and are obtained from the numerical solution. The solid line constructed simply to smoothly join the circles, is found to coincide precisely with the prediction of our analytic theory.}
  \label{fig2}
  \end{figure}

The result we found ( see Fig. \ref{fig2} ) is that the critical value
of the mask width $L$, i.e., the smallest value that can support a
non-zero bacterial population in the steady state, is lowered for a finite intermask distance $d$. We considered 10 different values of the hostile length (the intermask distance $d$) and varied the favorable length (the mask width $L$.) For each $L$-$d$ pair, we started with arbitrary initial conditions and let the program run until no time dependence was discernible.

We repeated the procedure for each of several sufficiently low values of $L$ and increased $L$ until the extinction disappeared. We also reversed the procedure starting with high values of $L$ and decreased them systematically until extinction appeared. Numerous runs allowed us to obtain corresponding pairs of $L$ and $d$ that mark the transition region. The results are denoted by filled circles in Fig. \ref{fig2}. The
shaded area represents the extinction region and the unshaded area the
parameter region in which bacterial population densities are non-zero in the steady state.
The curve passing through the numerically found
transition points may be considered, at this stage of our discussion, to be simply a smooth joining trace. We will see below that its exact shape can be accessed through our analytic theory.

It is easy to understand, on the basis of a qualitative argument, the shape and tendency of the results of the numerical solutions of the nonlinear equation. Bacteria diffuse from within the mask to the harsh region and die if they reach that region. Small values of the mask width or large values of the intermask distance result in extinction. The extinction effect is worsened by an increase in the intermask distance for small values of that distance but the effect saturates for larger values. Hence the saturation in the curve.

All these features can be tested experimentally in our proposed set-up. Quantitative comparison with the predictions of the Fisher equation are possible because we have developed an exact analytic theory of the interplay of the favorable and hostile distances as reflected in the transition. What makes the proposed comparison with experiment significant is that, although the Fisher equation, whose \emph{numerical} solution has led us to set out the separation curve between the extinction region and the rest in Fig. \ref{fig1}, cannot be solved exactly by analytic means, the separation curve itself can be obtained analytically. Indeed, we show below that the curve is given precisely by
\begin{equation}
L=\sqrt{\frac{D}{a}}\left[\arctan \sqrt {\frac{a_1}{a}}+\arctan \left[\sqrt{\frac{a_1}{a}} \tanh \left(d\sqrt{\frac{a_1}{D}}\right) \right]\right].
\label{analytic}
\end{equation}
This prediction, which is one of the central analytic results of the present paper, coincides with the solid curve in Fig. \ref{fig2}.

\section{Analytic Theory for the Twin Mask Set-up}
Following the ideas of Ludwig et al. \cite{ludwig} but applying them to the many-mask system, we consider the steady state of the Fisher equation (\ref{fisher}) and argue that if there is a transition, the quadratic term in the steady state $u(x)$ can be neglected at the extinction point in favor of the linear terms since $u(x)$ itself vanishes at the transition. We are thus led to seek the solutions of
\begin{equation}
\frac{d^2 u(x)}{d x^2}+\alpha^2 u(x)=0
\end{equation}
under the two masks, and of
\begin{equation}
\frac{d^2 u(x)}{d x^2}-\alpha_1^2 u(x)=0
\end{equation}
outside the masks, with $\alpha^2=a/D$ and $\alpha_1^2=a_1/D$. We take the masks to lie from $x=\pm d$ to $x=\pm (d + L)$. Using $I$ and $O$ as constants, the most general functions for $u(x)$ inside and outside the mask, respectively, are, as a result of the symmetry,
(we consider only the right side of the origin, since all considerations repeat unchanged on the left side by symmetry)
\begin{equation}
u(x)=I\cos (\alpha |x|-\phi)
\end{equation}
inside the mask, $d<x<d+L$,
\begin{equation}
u(x)=O \exp (-\alpha_1 |x|)
\end{equation}
in the extreme outside, in the harsh region, $x>d+L$, and
\begin{equation}
u(x)=C \cosh (\alpha_1 |x|)
\end{equation}
in the central region between the two masks, $-d<x<d$.
Matching the logarithmic derivative of the solution at the outer and the inner boundaries leads to
\begin{eqnarray}
\tan [\alpha (d+L)-\phi]&=&\alpha_1/\alpha \nonumber \\
\tan(\alpha d-\phi)&=&-(\alpha_1/\alpha)\tanh {\alpha_1 d}, \nonumber
\end{eqnarray}
and elimination of $\phi$ from these two equations leads to Eq. 
(\ref{analytic}) quoted above.
Notice how, in light of the behavior of the hyperbolic tangent, Eq. (\ref{analytic}), one of our central results, reduces to Ludwig et al.'s single mask value \cite{ludwig} of $L$, when $d$ attains infinite values, and half that value when $d$ vanishes. Both are fully expected and natural results. It is also instructive to rewrite the transition relation as
\begin{equation}
\mathcal{L}=\arctan \xi+\arctan (\xi \tanh {\delta\xi})
\label{twodimless}
\end{equation}
where we express the favorable distance $L$ and hostile distance $d$ normalized to 
the growth diffusion length $g=\sqrt{D/a}$ as $ \mathcal{L}=L/g$ and $\delta=d/g$ 
respectively, and the depletion parameter $\xi$ is the square root of the ratio 
of the destruction rate $a_1$ and the growth rate $a$.

\begin{figure}
 \centerline{\includegraphics[width=0.4\textwidth] {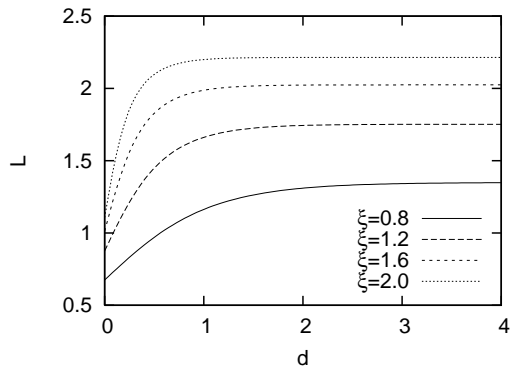}}
 \caption{ Our analytic predictions for the dependence of critical values of the favorable length $L$ and the hostile length $d$ respectively from Eq. (\ref{analytic}). As in other plots, the lengths are expressed as dimensionless ratios to the diffusion length $\sqrt{D/ a}$. Different curves correspond, as shown in the legend, to different values of $\xi$, i.e., to different intensities of the ultraviolet radiation. Note the doubling of $L$ as $d$ passes from $0$ to $\infty$ in all curves. }
\label{fig3}
 \end{figure}

We display in Fig. \ref{fig3} our analytic result, Eq. (\ref{analytic}), equivalently Eq. (\ref{twodimless}). As in other plots the lengths are expressed as dimensionless ratios to the diffusion length $\sqrt{D/a} $. Different curves correspond, as shown in the legend, to different values of $\xi$, i.e., to different intensities of the ultraviolet radiation. The curve for $\xi=3$ from our theory is shown as the solid line in Fig. \ref{fig2},  and displays exact coincidence with the numerical findings. To be noted is the fact that the numerical solutions are of the full nonlinear equation and have ben found for a specific $b$ whereas the analytic theory is linear and does not require the use of a value of $b$, requiring for its application only the condition $b>0$.

\begin{figure}
 \centering
 \centerline{\includegraphics[width=0.4\textwidth] {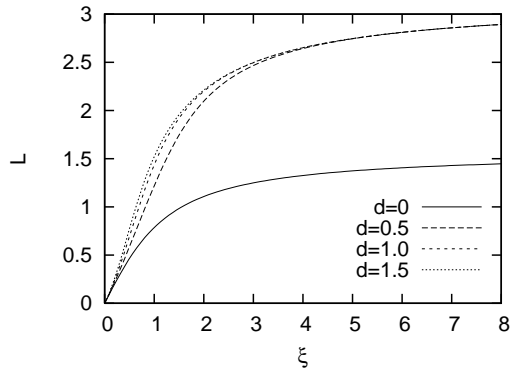}}
 \caption{Our analytical prediction from Eq. (\ref{analytic}) for the dependence of the critical value of $L$ on the intensity of ultraviolet radiation, represented here by the value of $\xi$, for given values of the hostile length (half the intermask separation) $d $ as shown. All $d$ values produce the same saturation values of $L$ for infinitely harsh ultraviolet intensity except for $d=0$: all curves reach the same higher value for large $\xi$ except the $d=0$ curve whose saturation value is one half that of the others. See text for explanation. The lengths are expressed as dimensionless ratios to the diffusion length $\sqrt{D/a}$.}
\label{fig4}
 \end{figure}

It is clear from our Eq. (\ref{analytic}) that the variation of the destructive rate $a_1$ may also be employed for useful experimental exploration, at least in principle. While this could have also been done in the one-mask scenario of ref. \cite{perry}, we explain the idea here in the two-mask case. Additional confirmation of the theoretical picture of the dynamics of the bacteria might be provided by observing how the critical $L$-$d$ values change with the \emph{intensity} of the ultraviolet light. Even if we do not know the precise dependence of $a_1$ on the intensity, we can be fairly certain that it increases with the latter, and undergoes a saturation at high values of the intensity. A variation of the critical value of $L$ with the intensity would show therefore qualitative behavior similar to the variation with $a_1$, equivalently with the ratio $\xi= \sqrt{a_1/a}$. The latter variation is displayed in Fig. \ref{fig4}. With one exception, all $d$ values produce the same saturation value of $L$ for large $\xi$. The exception is $d=0$: its saturation value is one half that of the others. Mathematically, this corresponds to the hyperbolic tangent becoming $1$ for all non-zero values of $d$ for large enough $ \xi$, but vanishing if $d$ vanishes. Physically, this means that if there is an adverse region between the masks, bacteria will be killed on arriving there, in light of the infinitely harsh radiation, reducing the problem to a single-mask scenario with the given $L$ as the mask width. On the other hand, if the intermediate harsh region does not exist at all (because $d=0$), one is reduced to considering a \emph{single} mask with width twice that of the given mask.

Our suggestion, thus, is to use both the relative variation of the favorable and hostile lengths, $L$ and $d$, and of the intensity of ultraviolet light in the manner discussed, to check our simple quantitative predictions.

\section{Multiple Mask Set-up}
It is possible and useful to construct set-ups with \emph{multiple} masks for further experimental verification.
The theory for this situation along the Ludwig et al. arguments we have developed here is slightly more tedious to write down. For ease in notation, let us adopt as we have done in Eq. (\ref{twodimless}), the symbols $\mathcal{L}$ and $\delta$ to express the favorable and hostile lengths in units of $\sqrt{D/a}$. Then the linear Petri dish formula for an even number of masks is obtained by executing the following pseudocode.

Let $H$ be an operator defined by the successive operations of (i) taking the hyperbolic tangent of what it acts on, (ii) multiplying the result by $\xi$, (iii) taking the arctangent of the result, and (iv) subtracting the result from $\mathcal{L}$. Define the operator $T$ similarly through the successive operations of (i) taking the trigonometric tangent of what it acts on, (ii) dividing the result by $ \xi$, (iii) taking the hyperbolic arctangent of the result, and (iv) subtracting the result from $2\xi \delta$.To obtain the required formula for an even number of masks, start with the product $\xi \delta $. Apply $H$ and then $T$ alternately and successively so that there are as many $H$'s in the operation as the number of mask pairs in the system, but one more $H$ than $T$. (Thus, do $H$ for a pair of masks, $HTH$ for 4 masks, $HTHTH$ for 6 masks, and so on.) Finally, equate the result to the arctangent of $\xi$ to get an implicit formula showing the $\mathcal{L}$-$\delta$ relation. The pseudocode can be expressed succinctly in terms of 
operators $T$ and $H$ by stating that the general expression for $2n$ masks can be formally written as $\left[ H\displaystyle\prod_{i=1}^{n-1}(TH)_{i}\right] \left(\xi\delta\right)=\arctan \xi$. As can be verified, this reproduces 
the 2-mask formula, Eq. (\ref{analytic}). The 4-mask formula is

\begin{eqnarray}
\mathcal{L} &=& \arctan \xi + \arctan\bigg(\xi\tanh\Big(2\xi\delta - \rm{arctanh}\big((1/\xi)\nonumber\\
 && \tan(\mathcal{L}-\arctan(\xi\tanh\,\xi\delta))\big)\Big)\bigg).
\end{eqnarray}

\section{Circular Petri Dish}
\begin{figure}
 \centering
 \centerline{\includegraphics[width=0.4\textwidth]{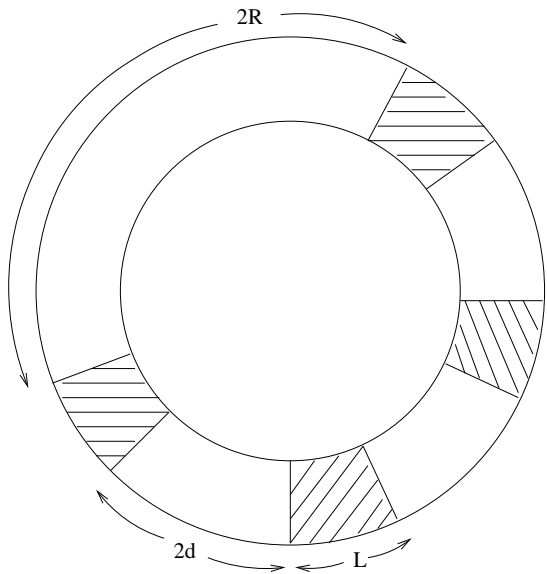}}
 \caption{ The proposed set-up in a \emph{circular} geometry shown here with a multiple mask set-up with 3   experimentally controllable parameters: $R$ as well as $L$ and $d$.}
\label{fig5}
 \end{figure}

Our suggested set-up and calculations have taken the Petri dish to have infinite extension away from the two masks, an assumption that should be reasonable in light of the destructive effects of the radiation. We now analyze another practical extension of the theory. Consider the two-mask situation in a \emph{circular} Petri dish, i.e. one in which the boundary conditions are periodic, the hostile distance between the edges of the masks being $2d$ on one side as in the earlier analysis but now also $2R$ on the other side, see Fig. \ref{fig5}. Let the width of the dish be small enough so that bacterial diffusion can be considered to be still 1-dimensional but over a total distance of extent $2(L+d +R)$. We are
not interested here in 2-dimensional considerations appropriate to wide dishes, an example of which may be found in ref.
\cite{shnerb2} for experiments with \emph{moving} masks or rotating dishes.

We now have, as in the analysis given above, $u(x)=I\cos (\alpha |x|- \phi)$ inside the right mask, i.e., for $d<x<d+L$, but to the right of the right mask, i.e., for $x>d+L$, $u(x)$ is not proportional to $ \exp (-\alpha_1 |x|)$ as in the earlier analysis, but
\begin{equation}
u(x)=O\cosh (\alpha_1 |x|-\eta).
\end{equation}
This expression, which, in the right region under consideration, where $x>0$, can be written as $u(x)=O\cosh (\alpha_1 x-\eta)$, will hold on the left extreme as well but there it simplifies to $u(x)=O\cosh (\alpha_1 x+\eta)$. The additional phase $\eta$ is immediately determined from the periodicity in the boundary conditions, i.e., by matching the logarithmic derivative of the solution at $|x|=d+L+R$:
\begin{equation}
\eta=\alpha_1(d+L+R).
\end{equation}
Matching also at $x=\pm d$ and $x=\pm (d+L)$, one arrives at the new formula for critical $L$ and $d$, valid for a \emph{circular} Petri dish:
\begin{eqnarray}\nonumber
L&=&\sqrt{\frac{D}{a}}\bigg(\arctan \Big(\sqrt {\frac{a_1}{a}}\tanh \big(R\sqrt{\frac{a_1}{D}}\big)\Big)+ \nonumber\\
 && + \arctan \Big(\sqrt{\frac{a_1}{a}} \tanh \big(d \sqrt{\frac{a_1}{D}}\big)\Big)\bigg).
\label{circular}
\end{eqnarray}

We see that the circular dish result, Eq. (\ref{circular}), is symmetric in the two hostile lengths $R$ and $d$ as it obviously must be, that it reduces to the result for the linear dish of infinite extent, viz., Eq. (\ref{analytic}), when $R$ becomes infinite since then the new hyperbolic tangent in Eq. (\ref{circular}) reduces to $1$, and that, when $R$ vanishes (equivalently when $d$ vanishes), it yields a new expression for the critical width for the single-mask case but with twice the width value. If there were simply a single mask in a circular Petri dish of extent $R$ and no $d$, the generalization of the Ludwig formula that would come out of our analysis would be
\begin{equation}
L=2 \sqrt{\frac{D}{a}}\arctan \left(\sqrt {\frac{a_1}{a}} \tanh(\alpha_1 R)\right).
\label{circludwig}
\end{equation}

Most importantly, we see that it should be possible in principle to use the new analytic result(s) for experimental probing of the phenomenon under consideration since one could construct and employ circular Petri dishes of desirable radius. Needless to say, the added effects discussed here would not be discernible for $R$ values that are very 
large as then bacteria would die in the external regions quickly enough.

We have also derived the corresponding expressions combining circular dish geometry with the multiple mask set-ups. Specifically, the 4-mask formula takes the form, $\mathcal{R}$ being $R$ measured in units of $\sqrt{D/a}$,
\begin{eqnarray}
\mathcal{L} &=& \arctan\big(\xi\tanh\xi\mathcal{R}\big) + \arctan \bigg(\xi\tanh\Big(2\xi\delta -\rm{arctanh}\big(\nonumber\\ &&(1/\xi)
 \tan(\mathcal{L}-\arctan(\xi\tanh\,\xi\delta))\big)\Big)\bigg).
\end{eqnarray}
whose reduction to the linear dish formula is obvious as $R$ becomes infinite. We have verified all these expressions that we have derived, by comparing their predictions to numerical solutions of the Fisher equation with explicit (arbitrary) non-zero $b$'s.
\section{Concluding Remarks}
The proposal for experimental observations of abrupt population transitions in patches that we have presented 
above should be of interest for multiple reasons. Patches or spatial inhomogeneities have been studied and emphasized early on in ecology by many authors such as Levin \cite{levin}, Murray \cite{murray} and Shigesada \cite{shigesada}
 in varied contexts including traveling waves, and there have been recent contributions as well on the Dirichlet conditions problem for general nonlinearity incorporating Allee effects \cite{mendez}.

Our analysis here provides a clean proposal for experiments which derives its potential for clear interpretation from the use of two (or more) controllable distances of travel, one favorable and the other unfavorable. One should be able to use it for extraction of parameter combinations in the Fisher equation which has been ubiquitous in mathematical ecology. The idea behind the present proposal should therefore find use in systems other than bacterial aggregates. Indeed, we have found similar abrupt transitions in \emph{infected} rodent populations in our study of the Hantavirus epidemic. The theory for that system is, however, more complex, and cannot be developed without approximation. By contrast, the theory we have presented here is, perhaps surprisingly, exact, i.e., analytical. This is so in spite of the fact that analytic solutions of the Fisher equation for these situations do not seem to be possible except as elliptic quadratures. The reason for this happy state of affairs can be understood from the original arguments of Ludwig et al. given in their 
paper on spatial patterning of the budworm \cite{ludwig}, or from the lucid explanations given, e.g., in a recent 
text on mathematical ecology \cite{kot}. At the transition, the densities vanish and therefore terms of order higher than the first may be safely neglected. Our theoretical contribution is only in generalizing that analysis to multiple masks and nonlinear geometries as shown. The  observationable controllable features our proposal emphasizes, and directly
utilizes, are (i) the interplay of the favorable and adverse lengths for bacterial traversal, (ii) the variation of 
the intensity of ultraviolet radiation which changes the interrelationship of the critical values, (iii) the multiple-mask set-up which reintroduces bacteria into favorable regions after they have passed into arid areas, and 
(iv) the optional use of circular geometry for the Petri dish which provides one more controllable length.

We are currently involved in investigating fluctuation effects that appear from detailed Monte Carlo considerations and become manifest for small population densities, in constructing the theory for transitions in partially infected populations  of rodents in open terrains, in analyzing the effects of static disorder in the placement and size of the masks, and in studying the effects of dynamic (time) variation of mask size and placement as in ref
\cite{ballard}. The bacterial experiments proposed above are, however, ready to go and we hope that observations will be made soon along the lines we have discussed.

\begin{acknowledgments}
This work was supported in part by the NSF under
grant no. INT-0336343, by NSF/NIH Ecology of Infectious Diseases under grant no. EF-0326757. One of us (VMK) thanks Gandhi M. Viswanathan for an insightful remark that helped the completion of this work.
\end{acknowledgments}


\begin{thebibliography}{}
\bibitem{phase} H. E. Stanley \emph{Introduction to Phase Transitions and Critical Phenomena}
               (Oxford University Press, 1971).
\bibitem{bifurc} S. H. Strogatz, \emph{Nonlinear Dynamics and Chaos: With Applications to Physics,
                Biology, Chemistry and Engineering} (Westview press, 1994).
\bibitem{popu} E. Renshaw, \emph{Modelling Biological Populations in Space and Time}
               (Cambridge University Press, 1991).
\bibitem{lin} A. L. Lin, B. Mann, G. Torres, B. Lincoln, J. Kas, and H. L. Swinney, Biophys. J.
             {\bf 87}, 75 (2004);
              see also A. L. Lin in \emph{Modern Challenges in Statistical Mechanics: Patterns, Growth, and the Interplay of Nonlinearity and Complexity,} eds. V. M. Kenkre and K. Lindenberg, AIP Conference Proceedings Volume 658 (ISBN 0-7354-0118-7, 2003).
\bibitem{nelson} D. R. Nelson and N. M. Shnerb, Phys. Rev. E {\bf 58}, 1383 (1998);
                 K. A. Dahmen, D. R. Nelson and N. M. Shnerb, J. Math. Biol. {\bf 41}, 1 (2000).
\bibitem{kk} V. M. Kenkre and M. N. Kuperman, Phys. Rev. E. {\bf 67}, 051921 (2003).
\bibitem{perry} N. Perry, J. R. Soc. Interface {\bf 2}, 379 (2005).
 \bibitem{other} J. Wakita, K. Komatsu, A. Nakahara, T. Matsuyama and M. Matsushita, J. Phys. Soc. Jpn. {\bf 63}, 1205 (1994).
\bibitem{mann} B. Mann, PhD thesis (Univ of Texas, Austin, 2001).
\bibitem{ballard} M. Ballard, V. M. Kenkre and M. N. Kuperman, Phys. Rev. E {\bf 70}, 031912 (2004).
\bibitem{general} H. C. Berg, Phys. Today {\bf 53}, 24 (2000);
                  E. Ben-Jacob, I. Cohen and H. Levine, Adv. Phys. {\bf 49}, 395 (2000).
\bibitem{fisher} R. A. Fisher, Ann. Eugen. London {\bf 7}, 355 (1937).
\bibitem{skellam} J. G. Skellam, Biometrika {\bf 38}, 196 (1951).
\bibitem{kiss} H. Kierstead and L. B. Slobodkin, J. Mar. Res. {\bf 12}, 141 (1953).
\bibitem{kot} M. Kot, \emph{Elements of Mathematical Ecology} (Cambridge University Press,
               Cambridge, 2003).
\bibitem{ludwig} D. Ludwig, D. G. Aronson and H. F. Weinberger, J. Math. Biol. {\bf 8}, 217 (1979).
\bibitem{shnerb2} N. M. Shnerb, Phys. Rev. E  {\bf 63}, 011906 (2000).
\bibitem{levin} S. A. Levin, Ann. Rev. Ecol. Syst. {\bf 7}, 287 (1976).
\bibitem{murray} J. D. Murray, \emph{Mathematical Biology}(Springer-Verlag, Berlin, 1993). 
\bibitem{shigesada} N. Kinezaki, K. Kawasaki, F. Takasu and N. Shigesada, Theor. Popul. Biol. {\bf 64}, 291 (2003);
                     N. Shigesada, K. Kawasaki and E. Teramoto, Theor. Popul. Biol. {\bf 30}, 143 (1986) ;
                     N. Shigesada and K. Kawasaki, \emph{Biological Invasions: Theory and Practice}
                     ( Oxford University Press, Oxford, 1997).
\bibitem{mendez} V. Mendez and D. Campos, Phys. Rev. E {\bf 77}, 022901 (2008).
\end{thebibliography}
\end{document}